\documentclass[aps,prl,twocolumn,floatfix,superscriptaddress]{revtex4-1}
\usepackage{graphicx}
\usepackage{amsmath}
\usepackage{txfonts}
\usepackage[colorlinks,citecolor=magenta,linkcolor=blue]{hyperref}

\begin{document}


\title{Combined Semilocal Exchange Potential with Dynamical Mean-Field Theory}


\author{Li Huang}
\email[Corresponding author: ]{lihuang.dmft@gmail.com}
\affiliation{Science and Technology on Surface Physics and Chemistry Laboratory, P.O. Box 9-35, Jiangyou 621908, China}
\author{Haiyan Lu}
\affiliation{Beijing National Laboratory for Condensed Matter Physics, and Institute of Physics, Chinese Academy of Sciences, Beijing 100190, China}
\affiliation{Science and Technology on Surface Physics and Chemistry Laboratory, P.O. Box 9-35, Jiangyou 621908, China}

\date{\today}


\begin{abstract}
The modern semilocal exchange potential is an accurate and efficient approximation to the exact exchange potential of density functional theory. We tried to combine it with the dynamical mean-field theory to derive a new first-principles many-body approach for studying correlated electronic materials. As a paradigm, this approach was employed to investigate the electronic structures and optical properties of strongly correlated ionic insulator YbS. Compared to the standard density functional theory plus dynamical mean-field theory which surprisingly failed to give an insulating solution, the new approach correctly captured all of the important characteristics of YbS. Not only an energy gap between a fully occupied Yb-4$f$ state and an unoccupied conduction band, but also an absence of Drude peak in the optical conductivity $\sigma(\omega)$ were successfully reproduced.
\end{abstract}

\maketitle


\emph{Introduction.} The strongly correlated systems, which exhibit many fascinating properties and unusual phenomena, have attracted numerous experimental and theoretical interests in last decades~\cite{RevModPhys.70.1039,jorge:2009}. There exists strong Coulomb interaction of $d$, $f$ electrons with each other and with itinerant electronic states of the materials. Classical band theory, such as the density functional theory which based on a single-particle or independent electron picture~\cite{PhysRev.140.A1133,PhysRev.136.B864}, works quite well for simple metals and semiconductors where the electron-electron interaction is weak. But it fails to give a correct description for the strongly correlated materials. To the best of our knowledge, so far the combination of density functional theory and single-site dynamical mean-field theory (DFT + DMFT) is probably the most powerful established method to study the electronic structures of strongly correlated materials~\cite{RevModPhys.78.865,RevModPhys.68.13}. In the framework of the DFT + DMFT method, the DFT part is responsible for providing a first-principles treatment for itinerant electrons, while the local interaction effects in localized electrons are tackled by the DMFT method in a non-perturbative many-body manner. Nowadays the DFT + DMFT method has been extensively employed to explore or explain the exotic physics in many strongly correlated materials, such as the (orbital-selective) Mott metal-insulator transitions and high-spin to low-spin transitions in transition metal oxides~\cite{PhysRevB.85.245110,PhysRevLett.102.146402,PhysRevB.82.195101}, charge dynamics and spin dynamics in iron-based unconventional superconductors~\cite{PhysRevLett.100.226402,yin:2011,yin:2014}, $4f$ localized-itinerant crossovers in rare-earth heavy-fermion compounds~\cite{Shim1615,PhysRevLett.108.016402,PhysRevB.94.075132}, and valence state fluctuations in actinides~\cite{shim:2007,PhysRevB.81.035105,Janoscheke:2015}.   


Many efforts have been devoted to improve and enhance the DFT + DMFT method in recent years. These improvements and enhancements can be roughly classified into two different aspects. One way is to adopt the cluster versions~\cite{RevModPhys.77.1027} or diagrammatic extensions~\cite{PhysRevB.75.045118,PhysRevB.77.033101,PhysRevB.88.115112} of DMFT to substitute for the single-site DMFT. For example, the cluster dynamical mean-field theory (CDMFT), the dynamical cluster approximation (DCA), and the dynamical vertex approximation (D$\Gamma$A) have been merged with the DFT to incorporate the non-local interaction effects and introduce momentum dependence to the electronic self-energy function~\cite{PhysRevLett.94.026404,PhysRevB.85.165103,toschi:2011}. Another means is to choose more powerful and accurate first-principles methods to provide a better starting point for the successive DMFT calculations. There are a few useful attempts in this respect, including the hybrid functional method (HYF)~\cite{jacob:2008}, the quasi-particle approximation (GWA)~\cite{PhysRevLett.90.086402}, and the screened exchange potential (SEx)~\cite{PhysRevLett.113.266403}.    

\begin{figure*}[t]
\centering
\includegraphics[width=\textwidth]{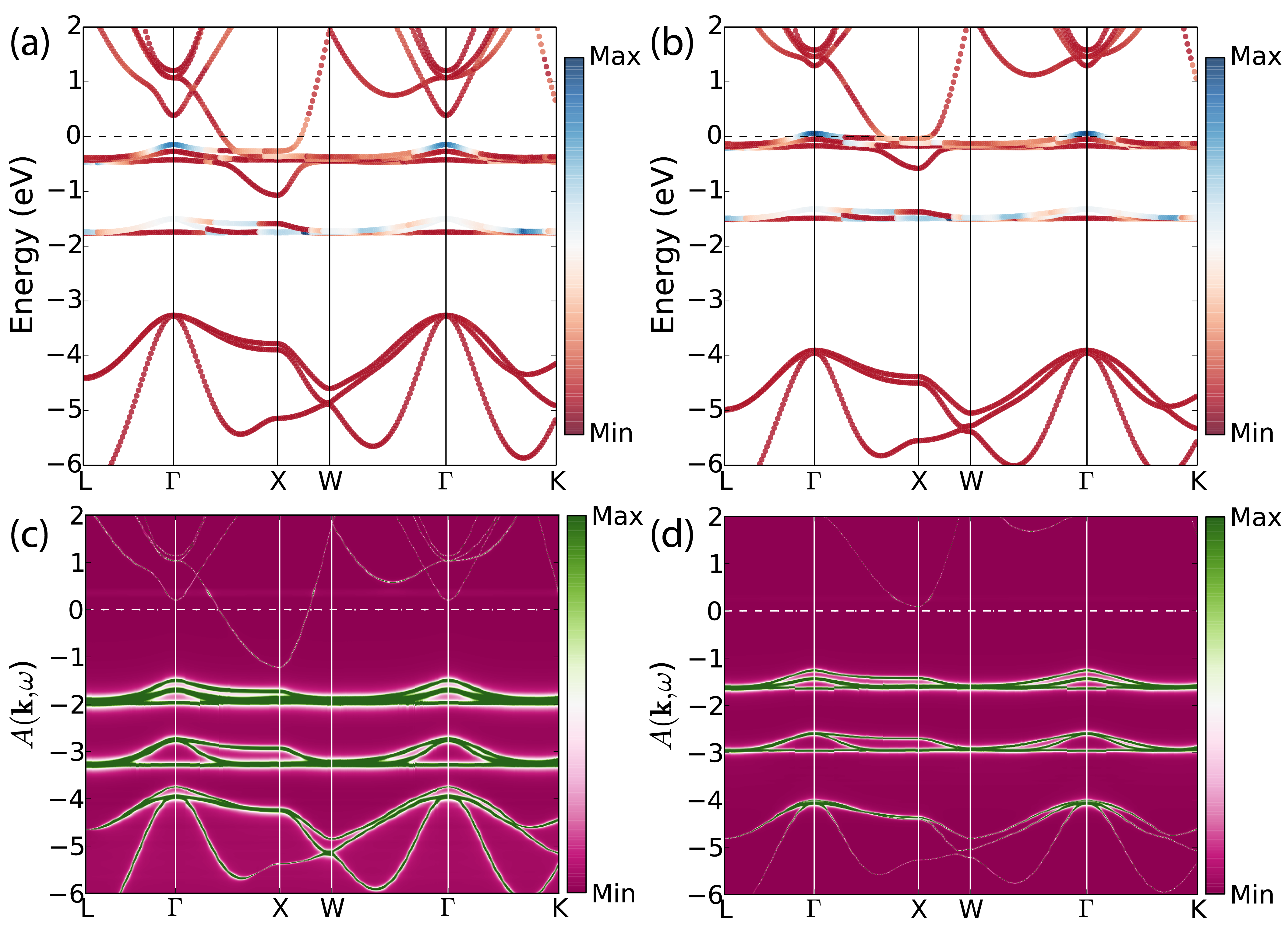}
\caption{(Color online). Electronic band structures of YbS under ambient pressure. (a) and (b) Fat bands obtained by the DFT + SOC and mBJ + SOC methods, respectively. The color bars denote the proportion of Yb-4$f$ character. (c) and (d) Momentum-resolved spectral functions $A(\mathbf{k},\omega)$ obtained by the DFT + DMFT and mBJ + DMFT methods, respectively. The color bars denote the spectral intensity. \label{fig:akw}}
\end{figure*}

\begin{figure*}[t]
\centering
\includegraphics[width=\textwidth]{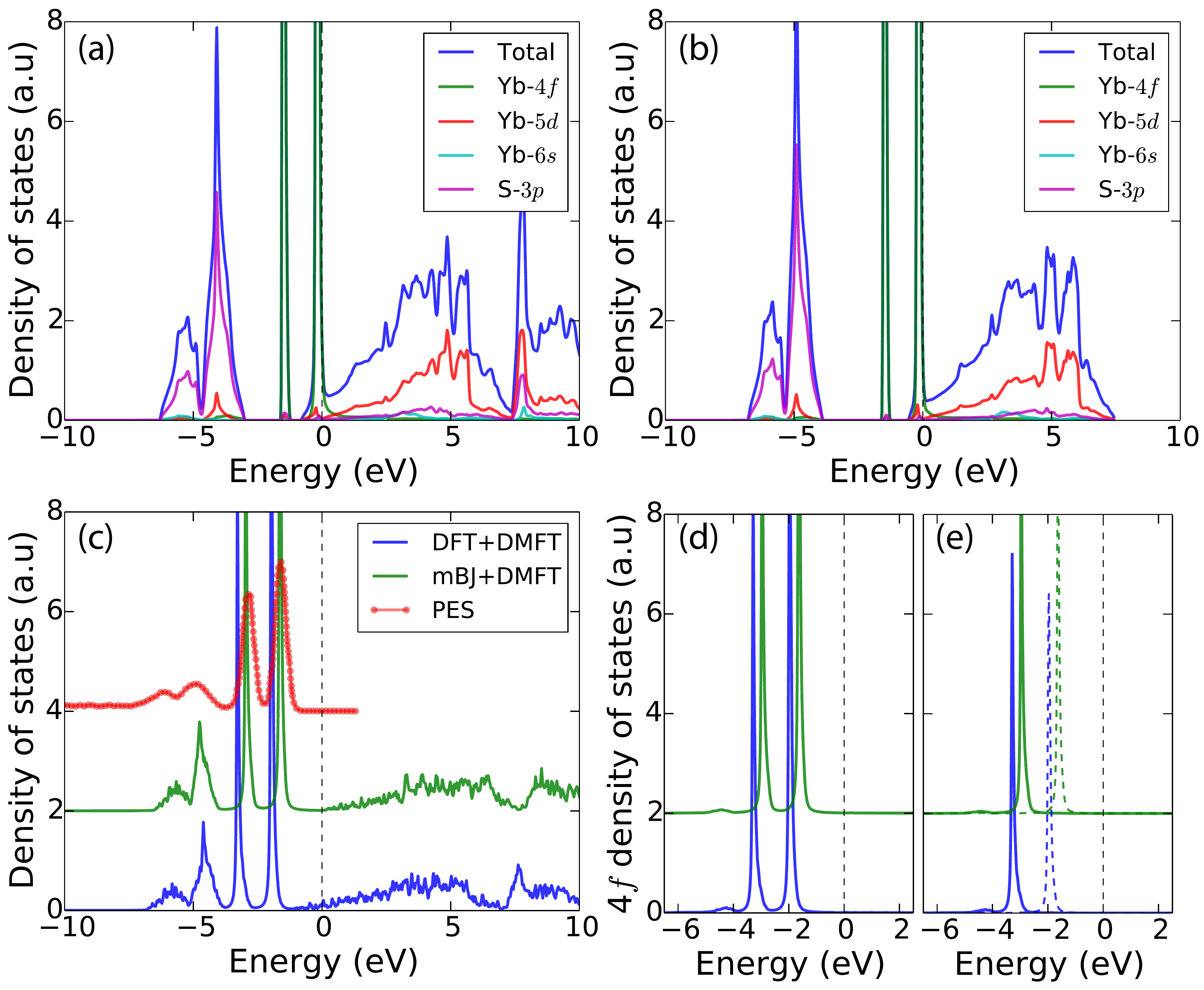}
\caption{(Color online). Total and partial density of states of YbS under ambient pressure. (a) and (b) Results obtained by the DFT + SOC and mBJ + SOC methods, respectively. (c) Results obtained by the DFT + DMFT and mBJ + DMFT methods. The experimental data (red filled circles) are extracted from Ref.~\cite{PhysRevB.78.195118}. (d) and (e) The Yb-$4f$ partial density of states by the mBJ + DMFT method. In (e) panel, the Yb $4f_{5/2}$ and $4f_{7/2}$ components are represented as solid and dashed lines, respectively. \label{fig:dos}}
\end{figure*}

\begin{figure*}[t]
\centering
\includegraphics[width=\textwidth]{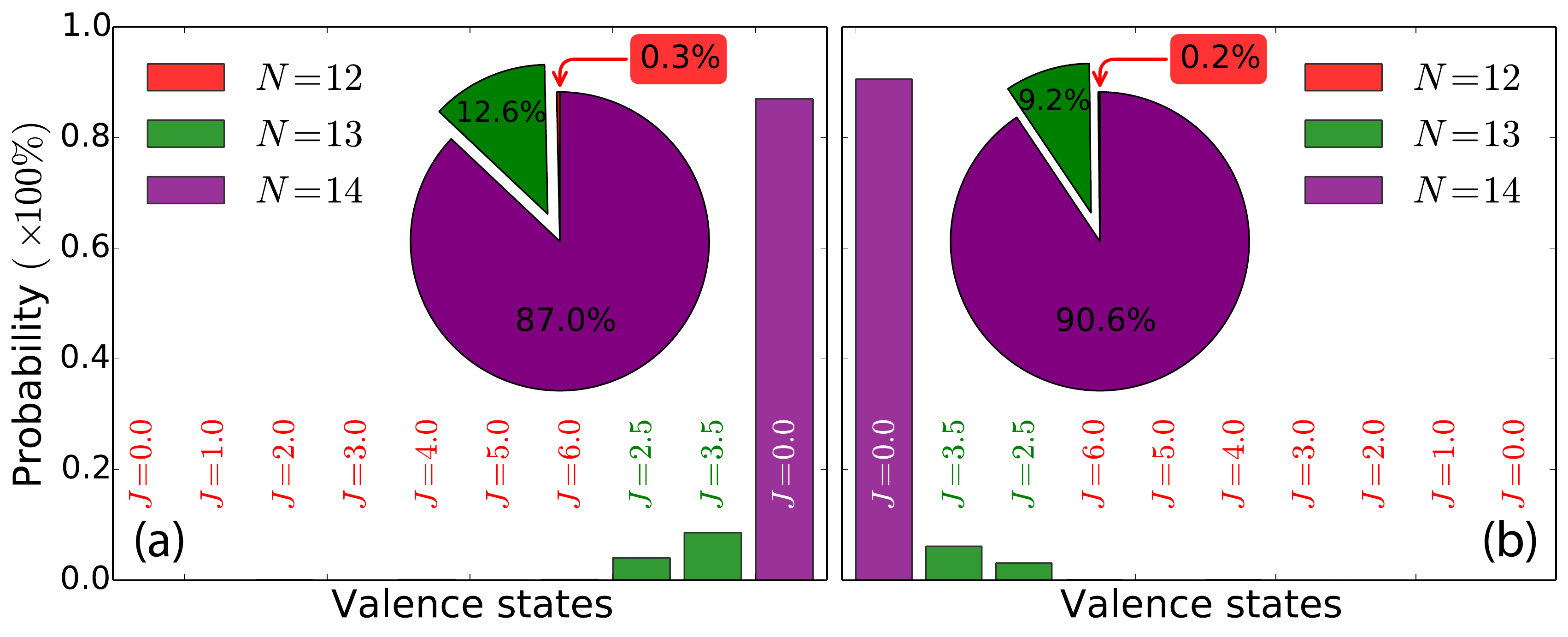}
\caption{(Color online). Valence state histograms of YbS under ambient pressure. The results are obtained by (a) the DFT + DMFT and (b) the mBJ + DMFT methods, respectively. The contributions from the $N = 12$ atomic eigenstates are too trivial to be seen in this figure. \label{fig:prob}}
\end{figure*}

\begin{figure}[t]
\centering
\includegraphics[width=\columnwidth]{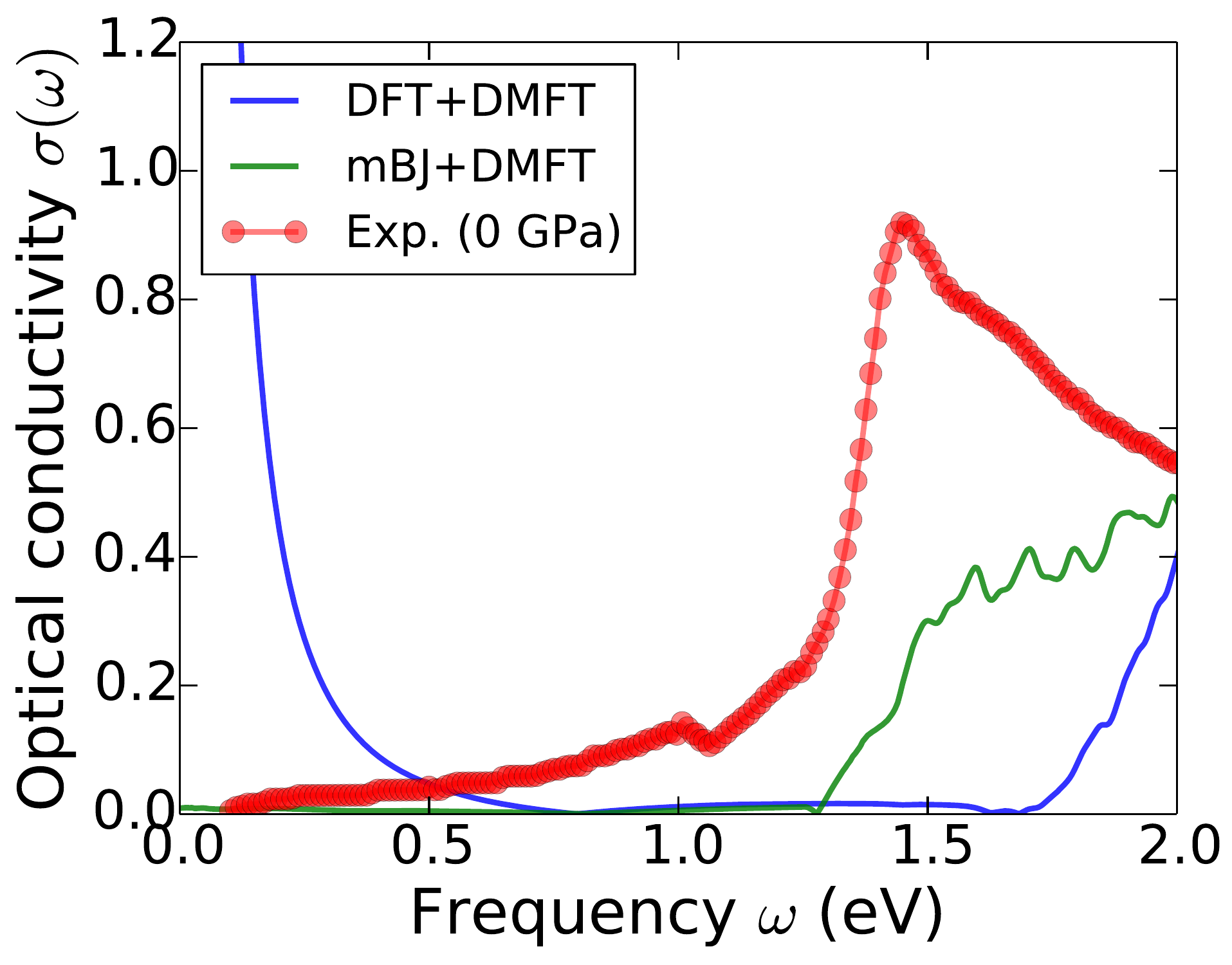}
\caption{(Color online). Real part of the optical conductivity $\Re \sigma(\omega)$. The experimental data (red filled circles) are taken from Ref.~\cite{PhysRevLett.103.237202}. \label{fig:optics}}
\end{figure}


In this paper we would like to propose a new first-principles many-body approach, namely, a combination of the state-of-the-art semilocal exchange potential (SEP) and the single-site DMFT. In the DFT, the electronic structures of materials are governed by the exchange-correlation potential $v_{\text{xc}}$, especially its exchange part $v_{\text{x}}$. The calculation of exact exchange potential (EXX) is highly nontrivial, and only possible by solving the optimized effective potential (OEP) equation~\cite{RevModPhys.80.3}. However, the SEP which depends on charge density ($\rho$), derivates of charge density ($\nabla \rho$ and $\nabla^2 \rho$), and kinetic energy density $\tau_s$, is rather simple and much faster than the EXX-OEP approach. So it has much broader applications in first-principles studies. In the standard DFT or DFT + DMFT calculations, the used SEPs usually base on local density approximation (LDA)~\cite{PhysRev.140.A1133} or generalized gradient approximation (GGA)~\cite{PhysRevLett.77.3865}. They barely resemble the EXX-OEP, so they generally tend to underestimate band gaps for semiconductors and make a coarse estimation for the impurity levels of strongly correlated materials. Recently, a new family of SEPs, i.e., the modified Becke-Johnson (mBJ) exchange potential and its variations were proposed by Tran \emph{et al.}~\cite{jcp_124_22,PhysRevLett.102.226401}, which mimic very well the behaviors of EXX-OEP~\cite{tran:2016}. Extensive tests on wide band gap insulators, $sp$ semiconductors, and strongly correlated $3d$ transition metal oxides etc., have verified their accuracy and usefulness. Motivated by these great achievements, here we try to combine the mBJ with the DMFT (mBJ + DMFT), and then apply it to explore the electronic structures of strongly correlated materials. The mBJ + DMFT scheme inherits the full merits of the mBJ exchange potential and can treat the energy levels of correlated and conducting bands correctly. Thus it is very suitable for studying correlated insulators. Furthermore, the mBJ + DMFT calculations are rarely more expensive than the standard DFT + DMFT calculations.


It is straightforward to perform the mBJ + DMFT calculations with the DFT + DMFT software package. The only modification is to active the mBJ exchange potential in the DFT part. Here we intend to use YbS as an example to demonstrate the accuracy of the mBJ + DMFT scheme. Under ambient pressure YbS crystallizes in a rock salt structure, and is a strongly correlated ionic insulator. Its band gap, which is consisted of fully occupied Yb-4$f$ state and an unoccupied conduction band, is about 1.30 eV~\cite{PhysRevB.78.195118,PhysRevB.32.8246}. We carried out the DFT, mBJ, DFT + DMFT, and mBJ + DMFT calculations, respectively, to uncover the electronic structures of YbS~\cite{how_to_calc}. The detailed results are presented and analyzed as follows. 


\emph{Momentum-resolved spectral functions.} The band structures $\epsilon_{n\mathbf{k}}$ and momentum-resolved spectral functions $A(\mathbf{k},\omega)$ of YbS are shown in Fig.~\ref{fig:akw}. Obviously, the band structures by the DFT and mBJ methods are metallic because there are a few bands crossing the Fermi level, which are in contrast to the experiments. Besides, there exists strong $c-f$ hybridization, especially near the $X$ point in the Brillouin zone. Note that with the mBJ exchange potential, the Yb-4$f$ bands are shifted toward the Fermi level slightly, while the other bands are pushed away. In the $A(\mathbf{k},\omega)$ obtained by the DFT + DMFT method, since the electronic correlation is already included, the Yb-$4f$ bands are shifted to high energy regime, and separated from the conduction bands completely. However, there are still bands in the Fermi level. In other words, the DFT + DMFT method fails to obtain an insulating state for YbS. Let us turn to the spectrum obtained by the mBJ + DMFT method. There is an indirect gap ($\sim 1.3$ eV) between the $\Gamma$ and $X$ points, which is in good agreement with the corresponding experiments~\cite{PhysRevB.78.195118,PhysRevB.32.8246}. Compared to the DFT + DMFT results, the Yb-4$f$ bands are shifted upward for $\sim 0.2$ eV.      

\emph{Electronic density of states.} Now let us focus on the total and partial density of states (DOS) of YbS (see Fig.~\ref{fig:dos}). Both the DFT and mBJ DOS show finite weights at the Fermi level, and a considerable hybridization develops between the Yb-4$f$ and Yb-$5d$ bands. The two sharp and intensive peaks near the Fermi level belong to the Yb $4f_{5/2}$ and $4f_{7/2}$ states, respectively, which are attributed to the splitting induced by the spin-orbital coupling (SOC) effect. The DOS by the DFT + DMFT method also shows small weights at the Fermi level which definitely indicates a metallic state. In addition, the peak positions for the Yb $4f_{5/2}$ and $4f_{7/2}$ states deviate from the experimental results significantly~\cite{PhysRevB.78.195118}. The DOS by the mBJ + DMFT method is consistent with the experimental data. Not only the band gap, but also the peaks for Yb-$4f$ and S-$3p$ bands are well reproduced. All this facts suggest that only the mBJ + DMFT method is capable of giving a correct picture of the electronic structures of YbS.  

\emph{Valence state fluctuations.} Valence instability is a key ingredient of the unusual properties of Yb-based materials~\cite{PhysRevLett.102.246401}. YbS is a typical mixed-valence compound~\cite{PhysRevLett.83.3900}. The nominal valence for Yb cation is +2, the corresponding Yb-4$f$ electronic configuration is $4f^{14}$. However, according to the experimental data by using the high-resolution X-ray absorption spectroscopy, the actual valence for Yb cation under ambient pressure is +2.08, i.e., the $4f$ occupancy is about 13.92~\cite{PhysRevB.87.115107}. Here, we used the DFT + DMFT and mBJ + DMFT methods to study the valence state fluctuations of YbS. The calculated results are presented in Fig.~\ref{fig:prob}. We find that no matter what method is used, the $|N = 14.0, J = 0.0\rangle$ atomic eigenstate ($4f^{14}$) is overwhelmingly dominant, which accounts for at least $> 87.0 \%$. The $N = 13$ atomic eigenstates ($4f^{13}$) is less important, which account for about 10\%. The probabilities for the $N = 12$ atomic eigenstates are trivial (about 0.2\%$ \sim $0.3\%) and can be ignored. Note that if the mBJ + DMFT method is used, the $|N = 14, J = 0.0\rangle$ atomic eigenstate has a higher probability, which implies the valence state fluctuation is less stronger than using the DFT + DMFT method. Besides, the mBJ + DMFT method predicts that $n_{4f} = 13.90$ and the valence is +2.10, which is somewhat closer to the experimental values~\cite{PhysRevB.87.115107} than the ones obtained by the DFT + DMFT method ($n_{4f} = 13.85$, and the valence is +2.15).

\emph{Optical properties.} The optical spectroscopy is a powerful tool to probe the electronic states of strongly correlated materials. We used the DFT + DMFT and mBJ + DMFT methods to calculate the optical conductivity $\sigma(\omega)$ of YbS, and then compared the calculated results with the available experimental data~\cite{PhysRevLett.103.237202}. The comparison is shown in Fig.~\ref{fig:optics}. The experimental $\sigma(\omega)$ of YbS exhibits prominent insulating characteristics. It is almost featureless when $\omega < 1.0$, but shows a ``dip" near 1.0 eV and a large peak around 1.5 eV~\cite{PhysRevLett.103.237202}. The optical conductivity obtained by the DFT + DMFT method manifests typical metallic-like feature, namely, a strong Drude peak at $\omega = 0$, which is mainly contributed from the Yb-$5d$ conduction band according to Fig.~\ref{fig:akw}(c). On the other hand, except for the magnitude of $\sigma(\omega)$, the calculated spectrum by the mBJ + DMFT method agrees quite well with the measured one. We believe that neglecting Hubbard interactions among the $spd$ conduction electrons in the mBJ + DMFT calculations might be responsible for this discrepancy~\cite{PhysRevLett.94.036401}.


\emph{Discussions.} In order to capture the insulating nature of YbS, the strong electronic correlations and SOC effects in Yb-4$f$ bands, and the energy levels for conduction bands (mainly Yb-$5d$ bands) have to be taken into accounts on the same footing. Apparently, the DFT method is given out at first. The mBJ method ignores the electronic correlation completely. The DFT + DMFT method considers the electronic correlation explicitly, but skips the effect of conduction bands and uses imprecise 4$f$ impurity levels to carry out calculations. So the two methods fail as well. In the mBJ + DMFT method, all these issues are treated carefully, so it wins. In that sense, the mBJ + DMFT method is better than the DFT + DMFT method. On the other hand, compared to the very time-consuming GWA + DMFT and HYF + DMFT methods, the mBJ + DMFT method has competitive accuracy, but its efficiency is as good as the standard DFT + DMFT method. Therefore it is also superior to the GWA + DMFT~\cite{PhysRevLett.90.086402} and HYF + DMFT~\cite{jacob:2008} methods.   

There are still many open and interesting questions to be answered. For example, under moderate pressure a transition from ionic insulator to heavy electron metal would occur in YbS~\cite{PhysRevLett.103.237202}. The valence state of Yb will be changed at the same time. Apparently, the DFT + DMFT method could not give a reasonable explanation for the transition. So a theoretical description for this transition is still lacking. It is highly promising to use the newly developed mBJ + DMFT method to solve this problem. On the other hand, the mBJ exchange potential has some adjustable parameters~\cite{PhysRevLett.102.226401} and variations~\cite{PhysRevB.91.165121}. Besides the mBJ exchange potential, there are also some other SEPs, such as the AK13~\cite{PhysRevLett.111.036402}, EV93~\cite{PhysRevB.47.13164} and LB94~\cite{PhysRevA.49.2421}. It should be very constructive if a comparable study for these SEPs is conducted. 


\emph{Conclusion.} In summary, we developed the mBJ + DMFT method by combining the mBJ exchange potential with the DMFT method. We then used it to study the electronic structures and optical properties of strongly correlated ionic insulator YbS. The calculated momentum-resolved spectral functions, density of states, atomic eigenstate probability, and optical conductivity are well consistent with the available experimental results. The insulating nature of YbS is properly reproduced. The DFT + DMFT method fails to get an insulating solution because of the inaccurate 4$f$ impurity levels and the improper treatment of conduction bands. 

\begin{acknowledgments}
This work was supported by the Natural Science Foundation of China (No.~11504340), Discipline Development Fund Project of Science and Technology on Surface Physics and Chemistry Laboratory (No.~201502), and Science Challenge Project of China.
\end{acknowledgments}


\bibliography{ybs}

\begin{thebibliography}{45}%
\makeatletter
\providecommand \@ifxundefined [1]{%
 \@ifx{#1\undefined}
}%
\providecommand \@ifnum [1]{%
 \ifnum #1\expandafter \@firstoftwo
 \else \expandafter \@secondoftwo
 \fi
}%
\providecommand \@ifx [1]{%
 \ifx #1\expandafter \@firstoftwo
 \else \expandafter \@secondoftwo
 \fi
}%
\providecommand \natexlab [1]{#1}%
\providecommand \enquote  [1]{``#1''}%
\providecommand \bibnamefont  [1]{#1}%
\providecommand \bibfnamefont [1]{#1}%
\providecommand \citenamefont [1]{#1}%
\providecommand \href@noop [0]{\@secondoftwo}%
\providecommand \href [0]{\begingroup \@sanitize@url \@href}%
\providecommand \@href[1]{\@@startlink{#1}\@@href}%
\providecommand \@@href[1]{\endgroup#1\@@endlink}%
\providecommand \@sanitize@url [0]{\catcode `\\12\catcode `\$12\catcode
  `\&12\catcode `\#12\catcode `\^12\catcode `\_12\catcode `\%12\relax}%
\providecommand \@@startlink[1]{}%
\providecommand \@@endlink[0]{}%
\providecommand \url  [0]{\begingroup\@sanitize@url \@url }%
\providecommand \@url [1]{\endgroup\@href {#1}{\urlprefix }}%
\providecommand \urlprefix  [0]{URL }%
\providecommand \Eprint [0]{\href }%
\providecommand \doibase [0]{http://dx.doi.org/}%
\providecommand \selectlanguage [0]{\@gobble}%
\providecommand \bibinfo  [0]{\@secondoftwo}%
\providecommand \bibfield  [0]{\@secondoftwo}%
\providecommand \translation [1]{[#1]}%
\providecommand \BibitemOpen [0]{}%
\providecommand \bibitemStop [0]{}%
\providecommand \bibitemNoStop [0]{.\EOS\space}%
\providecommand \EOS [0]{\spacefactor3000\relax}%
\providecommand \BibitemShut  [1]{\csname bibitem#1\endcsname}%
\let\auto@bib@innerbib\@empty
\bibitem [{\citenamefont {Imada}\ \emph {et~al.}(1998)\citenamefont {Imada},
  \citenamefont {Fujimori},\ and\ \citenamefont {Tokura}}]{RevModPhys.70.1039}%
  \BibitemOpen
  \bibfield  {author} {\bibinfo {author} {\bibfnamefont {M.}~\bibnamefont
  {Imada}}, \bibinfo {author} {\bibfnamefont {A.}~\bibnamefont {Fujimori}}, \
  and\ \bibinfo {author} {\bibfnamefont {Y.}~\bibnamefont {Tokura}},\ }\href
  {\doibase 10.1103/RevModPhys.70.1039} {\bibfield  {journal} {\bibinfo
  {journal} {Rev. Mod. Phys.}\ }\textbf {\bibinfo {volume} {70}},\ \bibinfo
  {pages} {1039} (\bibinfo {year} {1998})}\BibitemShut {NoStop}%
\bibitem [{\citenamefont {Quintanilla}\ and\ \citenamefont
  {Hooley}(2009)}]{jorge:2009}%
  \BibitemOpen
  \bibfield  {author} {\bibinfo {author} {\bibfnamefont {J.}~\bibnamefont
  {Quintanilla}}\ and\ \bibinfo {author} {\bibfnamefont {C.}~\bibnamefont
  {Hooley}},\ }\href {http://stacks.iop.org/2058-7058/22/i=06/a=38} {\bibfield
  {journal} {\bibinfo  {journal} {Physics World}\ }\textbf {\bibinfo {volume}
  {22}},\ \bibinfo {pages} {32} (\bibinfo {year} {2009})}\BibitemShut {NoStop}%
\bibitem [{\citenamefont {Kohn}\ and\ \citenamefont
  {Sham}(1965)}]{PhysRev.140.A1133}%
  \BibitemOpen
  \bibfield  {author} {\bibinfo {author} {\bibfnamefont {W.}~\bibnamefont
  {Kohn}}\ and\ \bibinfo {author} {\bibfnamefont {L.~J.}\ \bibnamefont
  {Sham}},\ }\href {\doibase 10.1103/PhysRev.140.A1133} {\bibfield  {journal}
  {\bibinfo  {journal} {Phys. Rev.}\ }\textbf {\bibinfo {volume} {140}},\
  \bibinfo {pages} {A1133} (\bibinfo {year} {1965})}\BibitemShut {NoStop}%
\bibitem [{\citenamefont {Hohenberg}\ and\ \citenamefont
  {Kohn}(1964)}]{PhysRev.136.B864}%
  \BibitemOpen
  \bibfield  {author} {\bibinfo {author} {\bibfnamefont {P.}~\bibnamefont
  {Hohenberg}}\ and\ \bibinfo {author} {\bibfnamefont {W.}~\bibnamefont
  {Kohn}},\ }\href {\doibase 10.1103/PhysRev.136.B864} {\bibfield  {journal}
  {\bibinfo  {journal} {Phys. Rev.}\ }\textbf {\bibinfo {volume} {136}},\
  \bibinfo {pages} {B864} (\bibinfo {year} {1964})}\BibitemShut {NoStop}%
\bibitem [{\citenamefont {Kotliar}\ \emph {et~al.}(2006)\citenamefont
  {Kotliar}, \citenamefont {Savrasov}, \citenamefont {Haule}, \citenamefont
  {Oudovenko}, \citenamefont {Parcollet},\ and\ \citenamefont
  {Marianetti}}]{RevModPhys.78.865}%
  \BibitemOpen
  \bibfield  {author} {\bibinfo {author} {\bibfnamefont {G.}~\bibnamefont
  {Kotliar}}, \bibinfo {author} {\bibfnamefont {S.~Y.}\ \bibnamefont
  {Savrasov}}, \bibinfo {author} {\bibfnamefont {K.}~\bibnamefont {Haule}},
  \bibinfo {author} {\bibfnamefont {V.~S.}\ \bibnamefont {Oudovenko}}, \bibinfo
  {author} {\bibfnamefont {O.}~\bibnamefont {Parcollet}}, \ and\ \bibinfo
  {author} {\bibfnamefont {C.~A.}\ \bibnamefont {Marianetti}},\ }\href
  {\doibase 10.1103/RevModPhys.78.865} {\bibfield  {journal} {\bibinfo
  {journal} {Rev. Mod. Phys.}\ }\textbf {\bibinfo {volume} {78}},\ \bibinfo
  {pages} {865} (\bibinfo {year} {2006})}\BibitemShut {NoStop}%
\bibitem [{\citenamefont {Georges}\ \emph {et~al.}(1996)\citenamefont
  {Georges}, \citenamefont {Kotliar}, \citenamefont {Krauth},\ and\
  \citenamefont {Rozenberg}}]{RevModPhys.68.13}%
  \BibitemOpen
  \bibfield  {author} {\bibinfo {author} {\bibfnamefont {A.}~\bibnamefont
  {Georges}}, \bibinfo {author} {\bibfnamefont {G.}~\bibnamefont {Kotliar}},
  \bibinfo {author} {\bibfnamefont {W.}~\bibnamefont {Krauth}}, \ and\ \bibinfo
  {author} {\bibfnamefont {M.~J.}\ \bibnamefont {Rozenberg}},\ }\href {\doibase
  10.1103/RevModPhys.68.13} {\bibfield  {journal} {\bibinfo  {journal} {Rev.
  Mod. Phys.}\ }\textbf {\bibinfo {volume} {68}},\ \bibinfo {pages} {13}
  (\bibinfo {year} {1996})}\BibitemShut {NoStop}%
\bibitem [{\citenamefont {Huang}\ \emph {et~al.}(2012)\citenamefont {Huang},
  \citenamefont {Wang},\ and\ \citenamefont {Dai}}]{PhysRevB.85.245110}%
  \BibitemOpen
  \bibfield  {author} {\bibinfo {author} {\bibfnamefont {L.}~\bibnamefont
  {Huang}}, \bibinfo {author} {\bibfnamefont {Y.}~\bibnamefont {Wang}}, \ and\
  \bibinfo {author} {\bibfnamefont {X.}~\bibnamefont {Dai}},\ }\href {\doibase
  10.1103/PhysRevB.85.245110} {\bibfield  {journal} {\bibinfo  {journal} {Phys.
  Rev. B}\ }\textbf {\bibinfo {volume} {85}},\ \bibinfo {pages} {245110}
  (\bibinfo {year} {2012})}\BibitemShut {NoStop}%
\bibitem [{\citenamefont {Kune\ifmmode~\check{s}\else \v{s}\fi{}}\ \emph
  {et~al.}(2009)\citenamefont {Kune\ifmmode~\check{s}\else \v{s}\fi{}},
  \citenamefont {Korotin}, \citenamefont {Korotin}, \citenamefont {Anisimov},\
  and\ \citenamefont {Werner}}]{PhysRevLett.102.146402}%
  \BibitemOpen
  \bibfield  {author} {\bibinfo {author} {\bibfnamefont {J.}~\bibnamefont
  {Kune\ifmmode~\check{s}\else \v{s}\fi{}}}, \bibinfo {author} {\bibfnamefont
  {D.~M.}\ \bibnamefont {Korotin}}, \bibinfo {author} {\bibfnamefont {M.~A.}\
  \bibnamefont {Korotin}}, \bibinfo {author} {\bibfnamefont {V.~I.}\
  \bibnamefont {Anisimov}}, \ and\ \bibinfo {author} {\bibfnamefont
  {P.}~\bibnamefont {Werner}},\ }\href {\doibase
  10.1103/PhysRevLett.102.146402} {\bibfield  {journal} {\bibinfo  {journal}
  {Phys. Rev. Lett.}\ }\textbf {\bibinfo {volume} {102}},\ \bibinfo {pages}
  {146402} (\bibinfo {year} {2009})}\BibitemShut {NoStop}%
\bibitem [{\citenamefont {Shorikov}\ \emph {et~al.}(2010)\citenamefont
  {Shorikov}, \citenamefont {Pchelkina}, \citenamefont {Anisimov},
  \citenamefont {Skornyakov},\ and\ \citenamefont
  {Korotin}}]{PhysRevB.82.195101}%
  \BibitemOpen
  \bibfield  {author} {\bibinfo {author} {\bibfnamefont {A.~O.}\ \bibnamefont
  {Shorikov}}, \bibinfo {author} {\bibfnamefont {Z.~V.}\ \bibnamefont
  {Pchelkina}}, \bibinfo {author} {\bibfnamefont {V.~I.}\ \bibnamefont
  {Anisimov}}, \bibinfo {author} {\bibfnamefont {S.~L.}\ \bibnamefont
  {Skornyakov}}, \ and\ \bibinfo {author} {\bibfnamefont {M.~A.}\ \bibnamefont
  {Korotin}},\ }\href {\doibase 10.1103/PhysRevB.82.195101} {\bibfield
  {journal} {\bibinfo  {journal} {Phys. Rev. B}\ }\textbf {\bibinfo {volume}
  {82}},\ \bibinfo {pages} {195101} (\bibinfo {year} {2010})}\BibitemShut
  {NoStop}%
\bibitem [{\citenamefont {Haule}\ \emph {et~al.}(2008)\citenamefont {Haule},
  \citenamefont {Shim},\ and\ \citenamefont
  {Kotliar}}]{PhysRevLett.100.226402}%
  \BibitemOpen
  \bibfield  {author} {\bibinfo {author} {\bibfnamefont {K.}~\bibnamefont
  {Haule}}, \bibinfo {author} {\bibfnamefont {J.~H.}\ \bibnamefont {Shim}}, \
  and\ \bibinfo {author} {\bibfnamefont {G.}~\bibnamefont {Kotliar}},\ }\href
  {\doibase 10.1103/PhysRevLett.100.226402} {\bibfield  {journal} {\bibinfo
  {journal} {Phys. Rev. Lett.}\ }\textbf {\bibinfo {volume} {100}},\ \bibinfo
  {pages} {226402} (\bibinfo {year} {2008})}\BibitemShut {NoStop}%
\bibitem [{\citenamefont {Yin}\ \emph {et~al.}(2011)\citenamefont {Yin},
  \citenamefont {Haule},\ and\ \citenamefont {Kotliar}}]{yin:2011}%
  \BibitemOpen
  \bibfield  {author} {\bibinfo {author} {\bibfnamefont {Z.~P.}\ \bibnamefont
  {Yin}}, \bibinfo {author} {\bibfnamefont {K.}~\bibnamefont {Haule}}, \ and\
  \bibinfo {author} {\bibfnamefont {G.}~\bibnamefont {Kotliar}},\ }\href
  {\doibase 10.1038/nphys1923} {\bibfield  {journal} {\bibinfo  {journal} {Nat
  Phys}\ }\textbf {\bibinfo {volume} {7}},\ \bibinfo {pages} {294} (\bibinfo
  {year} {2011})}\BibitemShut {NoStop}%
\bibitem [{\citenamefont {Yin}\ \emph {et~al.}(2014)\citenamefont {Yin},
  \citenamefont {Haule},\ and\ \citenamefont {Kotliar}}]{yin:2014}%
  \BibitemOpen
  \bibfield  {author} {\bibinfo {author} {\bibfnamefont {Z.~P.}\ \bibnamefont
  {Yin}}, \bibinfo {author} {\bibfnamefont {K.}~\bibnamefont {Haule}}, \ and\
  \bibinfo {author} {\bibfnamefont {G.}~\bibnamefont {Kotliar}},\ }\href
  {\doibase 10.1038/nphys3116} {\bibfield  {journal} {\bibinfo  {journal} {Nat
  Phys}\ }\textbf {\bibinfo {volume} {10}},\ \bibinfo {pages} {845} (\bibinfo
  {year} {2014})}\BibitemShut {NoStop}%
\bibitem [{\citenamefont {Shim}\ \emph
  {et~al.}(2007{\natexlab{a}})\citenamefont {Shim}, \citenamefont {Haule},\
  and\ \citenamefont {Kotliar}}]{Shim1615}%
  \BibitemOpen
  \bibfield  {author} {\bibinfo {author} {\bibfnamefont {J.~H.}\ \bibnamefont
  {Shim}}, \bibinfo {author} {\bibfnamefont {K.}~\bibnamefont {Haule}}, \ and\
  \bibinfo {author} {\bibfnamefont {G.}~\bibnamefont {Kotliar}},\ }\href
  {\doibase 10.1126/science.1149064} {\bibfield  {journal} {\bibinfo  {journal}
  {Science}\ }\textbf {\bibinfo {volume} {318}},\ \bibinfo {pages} {1615}
  (\bibinfo {year} {2007}{\natexlab{a}})}\BibitemShut {NoStop}%
\bibitem [{\citenamefont {Choi}\ \emph {et~al.}(2012)\citenamefont {Choi},
  \citenamefont {Min}, \citenamefont {Shim}, \citenamefont {Haule},\ and\
  \citenamefont {Kotliar}}]{PhysRevLett.108.016402}%
  \BibitemOpen
  \bibfield  {author} {\bibinfo {author} {\bibfnamefont {H.~C.}\ \bibnamefont
  {Choi}}, \bibinfo {author} {\bibfnamefont {B.~I.}\ \bibnamefont {Min}},
  \bibinfo {author} {\bibfnamefont {J.~H.}\ \bibnamefont {Shim}}, \bibinfo
  {author} {\bibfnamefont {K.}~\bibnamefont {Haule}}, \ and\ \bibinfo {author}
  {\bibfnamefont {G.}~\bibnamefont {Kotliar}},\ }\href {\doibase
  10.1103/PhysRevLett.108.016402} {\bibfield  {journal} {\bibinfo  {journal}
  {Phys. Rev. Lett.}\ }\textbf {\bibinfo {volume} {108}},\ \bibinfo {pages}
  {016402} (\bibinfo {year} {2012})}\BibitemShut {NoStop}%
\bibitem [{\citenamefont {Lu}\ and\ \citenamefont
  {Huang}(2016)}]{PhysRevB.94.075132}%
  \BibitemOpen
  \bibfield  {author} {\bibinfo {author} {\bibfnamefont {H.}~\bibnamefont
  {Lu}}\ and\ \bibinfo {author} {\bibfnamefont {L.}~\bibnamefont {Huang}},\
  }\href {\doibase 10.1103/PhysRevB.94.075132} {\bibfield  {journal} {\bibinfo
  {journal} {Phys. Rev. B}\ }\textbf {\bibinfo {volume} {94}},\ \bibinfo
  {pages} {075132} (\bibinfo {year} {2016})}\BibitemShut {NoStop}%
\bibitem [{\citenamefont {Shim}\ \emph
  {et~al.}(2007{\natexlab{b}})\citenamefont {Shim}, \citenamefont {Haule},\
  and\ \citenamefont {Kotliar}}]{shim:2007}%
  \BibitemOpen
  \bibfield  {author} {\bibinfo {author} {\bibfnamefont {J.~H.}\ \bibnamefont
  {Shim}}, \bibinfo {author} {\bibfnamefont {K.}~\bibnamefont {Haule}}, \ and\
  \bibinfo {author} {\bibfnamefont {G.}~\bibnamefont {Kotliar}},\ }\href
  {\doibase http://dx.doi.org/10.1038/nature05647} {\bibfield  {journal}
  {\bibinfo  {journal} {Nature}\ }\textbf {\bibinfo {volume} {446}},\ \bibinfo
  {pages} {513} (\bibinfo {year} {2007}{\natexlab{b}})}\BibitemShut {NoStop}%
\bibitem [{\citenamefont {Yee}\ \emph {et~al.}(2010)\citenamefont {Yee},
  \citenamefont {Kotliar},\ and\ \citenamefont {Haule}}]{PhysRevB.81.035105}%
  \BibitemOpen
  \bibfield  {author} {\bibinfo {author} {\bibfnamefont {C.-H.}\ \bibnamefont
  {Yee}}, \bibinfo {author} {\bibfnamefont {G.}~\bibnamefont {Kotliar}}, \ and\
  \bibinfo {author} {\bibfnamefont {K.}~\bibnamefont {Haule}},\ }\href
  {\doibase 10.1103/PhysRevB.81.035105} {\bibfield  {journal} {\bibinfo
  {journal} {Phys. Rev. B}\ }\textbf {\bibinfo {volume} {81}},\ \bibinfo
  {pages} {035105} (\bibinfo {year} {2010})}\BibitemShut {NoStop}%
\bibitem [{\citenamefont {Janoschek}\ \emph {et~al.}(2015)\citenamefont
  {Janoschek}, \citenamefont {Das}, \citenamefont {Chakrabarti}, \citenamefont
  {Abernathy}, \citenamefont {Lumsden}, \citenamefont {Lawrence}, \citenamefont
  {Thompson}, \citenamefont {Lander}, \citenamefont {Mitchell}, \citenamefont
  {Richmond}, \citenamefont {Ramos}, \citenamefont {Trouw}, \citenamefont
  {Zhu}, \citenamefont {Haule}, \citenamefont {Kotliar},\ and\ \citenamefont
  {Bauer}}]{Janoscheke:2015}%
  \BibitemOpen
  \bibfield  {author} {\bibinfo {author} {\bibfnamefont {M.}~\bibnamefont
  {Janoschek}}, \bibinfo {author} {\bibfnamefont {P.}~\bibnamefont {Das}},
  \bibinfo {author} {\bibfnamefont {B.}~\bibnamefont {Chakrabarti}}, \bibinfo
  {author} {\bibfnamefont {D.~L.}\ \bibnamefont {Abernathy}}, \bibinfo {author}
  {\bibfnamefont {M.~D.}\ \bibnamefont {Lumsden}}, \bibinfo {author}
  {\bibfnamefont {J.~M.}\ \bibnamefont {Lawrence}}, \bibinfo {author}
  {\bibfnamefont {J.~D.}\ \bibnamefont {Thompson}}, \bibinfo {author}
  {\bibfnamefont {G.~H.}\ \bibnamefont {Lander}}, \bibinfo {author}
  {\bibfnamefont {J.~N.}\ \bibnamefont {Mitchell}}, \bibinfo {author}
  {\bibfnamefont {S.}~\bibnamefont {Richmond}}, \bibinfo {author}
  {\bibfnamefont {M.}~\bibnamefont {Ramos}}, \bibinfo {author} {\bibfnamefont
  {F.}~\bibnamefont {Trouw}}, \bibinfo {author} {\bibfnamefont {J.-X.}\
  \bibnamefont {Zhu}}, \bibinfo {author} {\bibfnamefont {K.}~\bibnamefont
  {Haule}}, \bibinfo {author} {\bibfnamefont {G.}~\bibnamefont {Kotliar}}, \
  and\ \bibinfo {author} {\bibfnamefont {E.~D.}\ \bibnamefont {Bauer}},\ }\href
  {\doibase 10.1126/sciadv.1500188} {\bibfield  {journal} {\bibinfo  {journal}
  {Sci. Adv.}\ }\textbf {\bibinfo {volume} {1}},\ \bibinfo {pages} {1500188}
  (\bibinfo {year} {2015})}\BibitemShut {NoStop}%
\bibitem [{\citenamefont {Maier}\ \emph {et~al.}(2005)\citenamefont {Maier},
  \citenamefont {Jarrell}, \citenamefont {Pruschke},\ and\ \citenamefont
  {Hettler}}]{RevModPhys.77.1027}%
  \BibitemOpen
  \bibfield  {author} {\bibinfo {author} {\bibfnamefont {T.}~\bibnamefont
  {Maier}}, \bibinfo {author} {\bibfnamefont {M.}~\bibnamefont {Jarrell}},
  \bibinfo {author} {\bibfnamefont {T.}~\bibnamefont {Pruschke}}, \ and\
  \bibinfo {author} {\bibfnamefont {M.~H.}\ \bibnamefont {Hettler}},\ }\href
  {\doibase 10.1103/RevModPhys.77.1027} {\bibfield  {journal} {\bibinfo
  {journal} {Rev. Mod. Phys.}\ }\textbf {\bibinfo {volume} {77}},\ \bibinfo
  {pages} {1027} (\bibinfo {year} {2005})}\BibitemShut {NoStop}%
\bibitem [{\citenamefont {Toschi}\ \emph {et~al.}(2007)\citenamefont {Toschi},
  \citenamefont {Katanin},\ and\ \citenamefont {Held}}]{PhysRevB.75.045118}%
  \BibitemOpen
  \bibfield  {author} {\bibinfo {author} {\bibfnamefont {A.}~\bibnamefont
  {Toschi}}, \bibinfo {author} {\bibfnamefont {A.~A.}\ \bibnamefont {Katanin}},
  \ and\ \bibinfo {author} {\bibfnamefont {K.}~\bibnamefont {Held}},\ }\href
  {\doibase 10.1103/PhysRevB.75.045118} {\bibfield  {journal} {\bibinfo
  {journal} {Phys. Rev. B}\ }\textbf {\bibinfo {volume} {75}},\ \bibinfo
  {pages} {045118} (\bibinfo {year} {2007})}\BibitemShut {NoStop}%
\bibitem [{\citenamefont {Rubtsov}\ \emph {et~al.}(2008)\citenamefont
  {Rubtsov}, \citenamefont {Katsnelson},\ and\ \citenamefont
  {Lichtenstein}}]{PhysRevB.77.033101}%
  \BibitemOpen
  \bibfield  {author} {\bibinfo {author} {\bibfnamefont {A.~N.}\ \bibnamefont
  {Rubtsov}}, \bibinfo {author} {\bibfnamefont {M.~I.}\ \bibnamefont
  {Katsnelson}}, \ and\ \bibinfo {author} {\bibfnamefont {A.~I.}\ \bibnamefont
  {Lichtenstein}},\ }\href {\doibase 10.1103/PhysRevB.77.033101} {\bibfield
  {journal} {\bibinfo  {journal} {Phys. Rev. B}\ }\textbf {\bibinfo {volume}
  {77}},\ \bibinfo {pages} {033101} (\bibinfo {year} {2008})}\BibitemShut
  {NoStop}%
\bibitem [{\citenamefont {Rohringer}\ \emph {et~al.}(2013)\citenamefont
  {Rohringer}, \citenamefont {Toschi}, \citenamefont {Hafermann}, \citenamefont
  {Held}, \citenamefont {Anisimov},\ and\ \citenamefont
  {Katanin}}]{PhysRevB.88.115112}%
  \BibitemOpen
  \bibfield  {author} {\bibinfo {author} {\bibfnamefont {G.}~\bibnamefont
  {Rohringer}}, \bibinfo {author} {\bibfnamefont {A.}~\bibnamefont {Toschi}},
  \bibinfo {author} {\bibfnamefont {H.}~\bibnamefont {Hafermann}}, \bibinfo
  {author} {\bibfnamefont {K.}~\bibnamefont {Held}}, \bibinfo {author}
  {\bibfnamefont {V.~I.}\ \bibnamefont {Anisimov}}, \ and\ \bibinfo {author}
  {\bibfnamefont {A.~A.}\ \bibnamefont {Katanin}},\ }\href {\doibase
  10.1103/PhysRevB.88.115112} {\bibfield  {journal} {\bibinfo  {journal} {Phys.
  Rev. B}\ }\textbf {\bibinfo {volume} {88}},\ \bibinfo {pages} {115112}
  (\bibinfo {year} {2013})}\BibitemShut {NoStop}%
\bibitem [{\citenamefont {Biermann}\ \emph {et~al.}(2005)\citenamefont
  {Biermann}, \citenamefont {Poteryaev}, \citenamefont {Lichtenstein},\ and\
  \citenamefont {Georges}}]{PhysRevLett.94.026404}%
  \BibitemOpen
  \bibfield  {author} {\bibinfo {author} {\bibfnamefont {S.}~\bibnamefont
  {Biermann}}, \bibinfo {author} {\bibfnamefont {A.}~\bibnamefont {Poteryaev}},
  \bibinfo {author} {\bibfnamefont {A.~I.}\ \bibnamefont {Lichtenstein}}, \
  and\ \bibinfo {author} {\bibfnamefont {A.}~\bibnamefont {Georges}},\ }\href
  {\doibase 10.1103/PhysRevLett.94.026404} {\bibfield  {journal} {\bibinfo
  {journal} {Phys. Rev. Lett.}\ }\textbf {\bibinfo {volume} {94}},\ \bibinfo
  {pages} {026404} (\bibinfo {year} {2005})}\BibitemShut {NoStop}%
\bibitem [{\citenamefont {Lee}\ \emph {et~al.}(2012)\citenamefont {Lee},
  \citenamefont {Foyevtsova}, \citenamefont {Ferber}, \citenamefont {Aichhorn},
  \citenamefont {Jeschke},\ and\ \citenamefont
  {Valent\'{\i}}}]{PhysRevB.85.165103}%
  \BibitemOpen
  \bibfield  {author} {\bibinfo {author} {\bibfnamefont {H.}~\bibnamefont
  {Lee}}, \bibinfo {author} {\bibfnamefont {K.}~\bibnamefont {Foyevtsova}},
  \bibinfo {author} {\bibfnamefont {J.}~\bibnamefont {Ferber}}, \bibinfo
  {author} {\bibfnamefont {M.}~\bibnamefont {Aichhorn}}, \bibinfo {author}
  {\bibfnamefont {H.~O.}\ \bibnamefont {Jeschke}}, \ and\ \bibinfo {author}
  {\bibfnamefont {R.}~\bibnamefont {Valent\'{\i}}},\ }\href {\doibase
  10.1103/PhysRevB.85.165103} {\bibfield  {journal} {\bibinfo  {journal} {Phys.
  Rev. B}\ }\textbf {\bibinfo {volume} {85}},\ \bibinfo {pages} {165103}
  (\bibinfo {year} {2012})}\BibitemShut {NoStop}%
\bibitem [{\citenamefont {Toschi}\ \emph {et~al.}(2011)\citenamefont {Toschi},
  \citenamefont {Rohringer}, \citenamefont {Katanin},\ and\ \citenamefont
  {Held}}]{toschi:2011}%
  \BibitemOpen
  \bibfield  {author} {\bibinfo {author} {\bibfnamefont {A.}~\bibnamefont
  {Toschi}}, \bibinfo {author} {\bibfnamefont {G.}~\bibnamefont {Rohringer}},
  \bibinfo {author} {\bibfnamefont {A.}~\bibnamefont {Katanin}}, \ and\
  \bibinfo {author} {\bibfnamefont {K.}~\bibnamefont {Held}},\ }\href {\doibase
  10.1002/andp.201100036} {\bibfield  {journal} {\bibinfo  {journal} {Annalen
  der Physik}\ }\textbf {\bibinfo {volume} {523}},\ \bibinfo {pages} {698}
  (\bibinfo {year} {2011})}\BibitemShut {NoStop}%
\bibitem [{\citenamefont {Jacob}\ \emph {et~al.}(2008)\citenamefont {Jacob},
  \citenamefont {Haule},\ and\ \citenamefont {Kotliar}}]{jacob:2008}%
  \BibitemOpen
  \bibfield  {author} {\bibinfo {author} {\bibfnamefont {D.}~\bibnamefont
  {Jacob}}, \bibinfo {author} {\bibfnamefont {K.}~\bibnamefont {Haule}}, \ and\
  \bibinfo {author} {\bibfnamefont {G.}~\bibnamefont {Kotliar}},\ }\href
  {http://stacks.iop.org/0295-5075/84/i=5/a=57009} {\bibfield  {journal}
  {\bibinfo  {journal} {EPL}\ }\textbf {\bibinfo {volume} {84}},\ \bibinfo
  {pages} {57009} (\bibinfo {year} {2008})}\BibitemShut {NoStop}%
\bibitem [{\citenamefont {Biermann}\ \emph {et~al.}(2003)\citenamefont
  {Biermann}, \citenamefont {Aryasetiawan},\ and\ \citenamefont
  {Georges}}]{PhysRevLett.90.086402}%
  \BibitemOpen
  \bibfield  {author} {\bibinfo {author} {\bibfnamefont {S.}~\bibnamefont
  {Biermann}}, \bibinfo {author} {\bibfnamefont {F.}~\bibnamefont
  {Aryasetiawan}}, \ and\ \bibinfo {author} {\bibfnamefont {A.}~\bibnamefont
  {Georges}},\ }\href {\doibase 10.1103/PhysRevLett.90.086402} {\bibfield
  {journal} {\bibinfo  {journal} {Phys. Rev. Lett.}\ }\textbf {\bibinfo
  {volume} {90}},\ \bibinfo {pages} {086402} (\bibinfo {year}
  {2003})}\BibitemShut {NoStop}%
\bibitem [{\citenamefont {van Roekeghem}\ \emph {et~al.}(2014)\citenamefont
  {van Roekeghem}, \citenamefont {Ayral}, \citenamefont {Tomczak},
  \citenamefont {Casula}, \citenamefont {Xu}, \citenamefont {Ding},
  \citenamefont {Ferrero}, \citenamefont {Parcollet}, \citenamefont {Jiang},\
  and\ \citenamefont {Biermann}}]{PhysRevLett.113.266403}%
  \BibitemOpen
  \bibfield  {author} {\bibinfo {author} {\bibfnamefont {A.}~\bibnamefont {van
  Roekeghem}}, \bibinfo {author} {\bibfnamefont {T.}~\bibnamefont {Ayral}},
  \bibinfo {author} {\bibfnamefont {J.~M.}\ \bibnamefont {Tomczak}}, \bibinfo
  {author} {\bibfnamefont {M.}~\bibnamefont {Casula}}, \bibinfo {author}
  {\bibfnamefont {N.}~\bibnamefont {Xu}}, \bibinfo {author} {\bibfnamefont
  {H.}~\bibnamefont {Ding}}, \bibinfo {author} {\bibfnamefont {M.}~\bibnamefont
  {Ferrero}}, \bibinfo {author} {\bibfnamefont {O.}~\bibnamefont {Parcollet}},
  \bibinfo {author} {\bibfnamefont {H.}~\bibnamefont {Jiang}}, \ and\ \bibinfo
  {author} {\bibfnamefont {S.}~\bibnamefont {Biermann}},\ }\href {\doibase
  10.1103/PhysRevLett.113.266403} {\bibfield  {journal} {\bibinfo  {journal}
  {Phys. Rev. Lett.}\ }\textbf {\bibinfo {volume} {113}},\ \bibinfo {pages}
  {266403} (\bibinfo {year} {2014})}\BibitemShut {NoStop}%
\bibitem [{\citenamefont {Matsunami}\ \emph {et~al.}(2008)\citenamefont
  {Matsunami}, \citenamefont {Chainani}, \citenamefont {Taguchi}, \citenamefont
  {Eguchi}, \citenamefont {Ishida}, \citenamefont {Takata}, \citenamefont
  {Okamura}, \citenamefont {Nanba}, \citenamefont {Yabashi}, \citenamefont
  {Tamasaku}, \citenamefont {Nishino}, \citenamefont {Ishikawa}, \citenamefont
  {Senba}, \citenamefont {Ohashi}, \citenamefont {Tsujii}, \citenamefont
  {Ochiai},\ and\ \citenamefont {Shin}}]{PhysRevB.78.195118}%
  \BibitemOpen
  \bibfield  {author} {\bibinfo {author} {\bibfnamefont {M.}~\bibnamefont
  {Matsunami}}, \bibinfo {author} {\bibfnamefont {A.}~\bibnamefont {Chainani}},
  \bibinfo {author} {\bibfnamefont {M.}~\bibnamefont {Taguchi}}, \bibinfo
  {author} {\bibfnamefont {R.}~\bibnamefont {Eguchi}}, \bibinfo {author}
  {\bibfnamefont {Y.}~\bibnamefont {Ishida}}, \bibinfo {author} {\bibfnamefont
  {Y.}~\bibnamefont {Takata}}, \bibinfo {author} {\bibfnamefont
  {H.}~\bibnamefont {Okamura}}, \bibinfo {author} {\bibfnamefont
  {T.}~\bibnamefont {Nanba}}, \bibinfo {author} {\bibfnamefont
  {M.}~\bibnamefont {Yabashi}}, \bibinfo {author} {\bibfnamefont
  {K.}~\bibnamefont {Tamasaku}}, \bibinfo {author} {\bibfnamefont
  {Y.}~\bibnamefont {Nishino}}, \bibinfo {author} {\bibfnamefont
  {T.}~\bibnamefont {Ishikawa}}, \bibinfo {author} {\bibfnamefont
  {Y.}~\bibnamefont {Senba}}, \bibinfo {author} {\bibfnamefont
  {H.}~\bibnamefont {Ohashi}}, \bibinfo {author} {\bibfnamefont
  {N.}~\bibnamefont {Tsujii}}, \bibinfo {author} {\bibfnamefont
  {A.}~\bibnamefont {Ochiai}}, \ and\ \bibinfo {author} {\bibfnamefont
  {S.}~\bibnamefont {Shin}},\ }\href {\doibase 10.1103/PhysRevB.78.195118}
  {\bibfield  {journal} {\bibinfo  {journal} {Phys. Rev. B}\ }\textbf {\bibinfo
  {volume} {78}},\ \bibinfo {pages} {195118} (\bibinfo {year}
  {2008})}\BibitemShut {NoStop}%
\bibitem [{\citenamefont {Matsunami}\ \emph {et~al.}(2009)\citenamefont
  {Matsunami}, \citenamefont {Okamura}, \citenamefont {Ochiai},\ and\
  \citenamefont {Nanba}}]{PhysRevLett.103.237202}%
  \BibitemOpen
  \bibfield  {author} {\bibinfo {author} {\bibfnamefont {M.}~\bibnamefont
  {Matsunami}}, \bibinfo {author} {\bibfnamefont {H.}~\bibnamefont {Okamura}},
  \bibinfo {author} {\bibfnamefont {A.}~\bibnamefont {Ochiai}}, \ and\ \bibinfo
  {author} {\bibfnamefont {T.}~\bibnamefont {Nanba}},\ }\href {\doibase
  10.1103/PhysRevLett.103.237202} {\bibfield  {journal} {\bibinfo  {journal}
  {Phys. Rev. Lett.}\ }\textbf {\bibinfo {volume} {103}},\ \bibinfo {pages}
  {237202} (\bibinfo {year} {2009})}\BibitemShut {NoStop}%
\bibitem [{\citenamefont {K\"ummel}\ and\ \citenamefont
  {Kronik}(2008)}]{RevModPhys.80.3}%
  \BibitemOpen
  \bibfield  {author} {\bibinfo {author} {\bibfnamefont {S.}~\bibnamefont
  {K\"ummel}}\ and\ \bibinfo {author} {\bibfnamefont {L.}~\bibnamefont
  {Kronik}},\ }\href {\doibase 10.1103/RevModPhys.80.3} {\bibfield  {journal}
  {\bibinfo  {journal} {Rev. Mod. Phys.}\ }\textbf {\bibinfo {volume} {80}},\
  \bibinfo {pages} {3} (\bibinfo {year} {2008})}\BibitemShut {NoStop}%
\bibitem [{\citenamefont {Perdew}\ \emph {et~al.}(1996)\citenamefont {Perdew},
  \citenamefont {Burke},\ and\ \citenamefont
  {Ernzerhof}}]{PhysRevLett.77.3865}%
  \BibitemOpen
  \bibfield  {author} {\bibinfo {author} {\bibfnamefont {J.~P.}\ \bibnamefont
  {Perdew}}, \bibinfo {author} {\bibfnamefont {K.}~\bibnamefont {Burke}}, \
  and\ \bibinfo {author} {\bibfnamefont {M.}~\bibnamefont {Ernzerhof}},\ }\href
  {\doibase 10.1103/PhysRevLett.77.3865} {\bibfield  {journal} {\bibinfo
  {journal} {Phys. Rev. Lett.}\ }\textbf {\bibinfo {volume} {77}},\ \bibinfo
  {pages} {3865} (\bibinfo {year} {1996})}\BibitemShut {NoStop}%
\bibitem [{\citenamefont {Becke}\ and\ \citenamefont
  {Johnson}(2006)}]{jcp_124_22}%
  \BibitemOpen
  \bibfield  {author} {\bibinfo {author} {\bibfnamefont {A.~D.}\ \bibnamefont
  {Becke}}\ and\ \bibinfo {author} {\bibfnamefont {E.~R.}\ \bibnamefont
  {Johnson}},\ }\href {\doibase http://dx.doi.org/10.1063/1.2213970} {\bibfield
   {journal} {\bibinfo  {journal} {J. Chem. Phys.}\ }\textbf {\bibinfo {volume}
  {124}},\ \bibinfo {pages} {221101} (\bibinfo {year} {2006})}\BibitemShut
  {NoStop}%
\bibitem [{\citenamefont {Tran}\ and\ \citenamefont
  {Blaha}(2009)}]{PhysRevLett.102.226401}%
  \BibitemOpen
  \bibfield  {author} {\bibinfo {author} {\bibfnamefont {F.}~\bibnamefont
  {Tran}}\ and\ \bibinfo {author} {\bibfnamefont {P.}~\bibnamefont {Blaha}},\
  }\href {\doibase 10.1103/PhysRevLett.102.226401} {\bibfield  {journal}
  {\bibinfo  {journal} {Phys. Rev. Lett.}\ }\textbf {\bibinfo {volume} {102}},\
  \bibinfo {pages} {226401} (\bibinfo {year} {2009})}\BibitemShut {NoStop}%
\bibitem [{\citenamefont {Tran}\ \emph {et~al.}(2016)\citenamefont {Tran},
  \citenamefont {Blaha}, \citenamefont {Betzinger},\ and\ \citenamefont
  {Blügel}}]{tran:2016}%
  \BibitemOpen
  \bibfield  {author} {\bibinfo {author} {\bibfnamefont {F.}~\bibnamefont
  {Tran}}, \bibinfo {author} {\bibfnamefont {P.}~\bibnamefont {Blaha}},
  \bibinfo {author} {\bibfnamefont {M.}~\bibnamefont {Betzinger}}, \ and\
  \bibinfo {author} {\bibfnamefont {S.}~\bibnamefont {Blügel}},\ }\href@noop
  {} {\  (\bibinfo {year} {2016})},\ \Eprint {http://arxiv.org/abs/1608.08415}
  {arXiv:1608.08415 [cond-mat]} \BibitemShut {NoStop}%
\bibitem [{\citenamefont {Syassen}\ \emph {et~al.}(1985)\citenamefont
  {Syassen}, \citenamefont {Winzen}, \citenamefont {Zimmer}, \citenamefont
  {Tups},\ and\ \citenamefont {Leger}}]{PhysRevB.32.8246}%
  \BibitemOpen
  \bibfield  {author} {\bibinfo {author} {\bibfnamefont {K.}~\bibnamefont
  {Syassen}}, \bibinfo {author} {\bibfnamefont {H.}~\bibnamefont {Winzen}},
  \bibinfo {author} {\bibfnamefont {H.~G.}\ \bibnamefont {Zimmer}}, \bibinfo
  {author} {\bibfnamefont {H.}~\bibnamefont {Tups}}, \ and\ \bibinfo {author}
  {\bibfnamefont {J.~M.}\ \bibnamefont {Leger}},\ }\href {\doibase
  10.1103/PhysRevB.32.8246} {\bibfield  {journal} {\bibinfo  {journal} {Phys.
  Rev. B}\ }\textbf {\bibinfo {volume} {32}},\ \bibinfo {pages} {8246}
  (\bibinfo {year} {1985})}\BibitemShut {NoStop}%
\bibitem [{how()}]{how_to_calc}%
  \BibitemOpen
  \href@noop {} {}\bibinfo {note} {See Supplementary Materials for
  computational details.}\BibitemShut {Stop}%
\bibitem [{\citenamefont {Ylvisaker}\ \emph {et~al.}(2009)\citenamefont
  {Ylvisaker}, \citenamefont {Kune\ifmmode~\check{s}\else \v{s}\fi{}},
  \citenamefont {McMahan},\ and\ \citenamefont
  {Pickett}}]{PhysRevLett.102.246401}%
  \BibitemOpen
  \bibfield  {author} {\bibinfo {author} {\bibfnamefont {E.~R.}\ \bibnamefont
  {Ylvisaker}}, \bibinfo {author} {\bibfnamefont {J.}~\bibnamefont
  {Kune\ifmmode~\check{s}\else \v{s}\fi{}}}, \bibinfo {author} {\bibfnamefont
  {A.~K.}\ \bibnamefont {McMahan}}, \ and\ \bibinfo {author} {\bibfnamefont
  {W.~E.}\ \bibnamefont {Pickett}},\ }\href {\doibase
  10.1103/PhysRevLett.102.246401} {\bibfield  {journal} {\bibinfo  {journal}
  {Phys. Rev. Lett.}\ }\textbf {\bibinfo {volume} {102}},\ \bibinfo {pages}
  {246401} (\bibinfo {year} {2009})}\BibitemShut {NoStop}%
\bibitem [{\citenamefont {Temmerman}\ \emph {et~al.}(1999)\citenamefont
  {Temmerman}, \citenamefont {Szotek}, \citenamefont {Svane}, \citenamefont
  {Strange}, \citenamefont {Winter}, \citenamefont {Delin}, \citenamefont
  {Johansson}, \citenamefont {Eriksson}, \citenamefont {Fast},\ and\
  \citenamefont {Wills}}]{PhysRevLett.83.3900}%
  \BibitemOpen
  \bibfield  {author} {\bibinfo {author} {\bibfnamefont {W.~M.}\ \bibnamefont
  {Temmerman}}, \bibinfo {author} {\bibfnamefont {Z.}~\bibnamefont {Szotek}},
  \bibinfo {author} {\bibfnamefont {A.}~\bibnamefont {Svane}}, \bibinfo
  {author} {\bibfnamefont {P.}~\bibnamefont {Strange}}, \bibinfo {author}
  {\bibfnamefont {H.}~\bibnamefont {Winter}}, \bibinfo {author} {\bibfnamefont
  {A.}~\bibnamefont {Delin}}, \bibinfo {author} {\bibfnamefont
  {B.}~\bibnamefont {Johansson}}, \bibinfo {author} {\bibfnamefont
  {O.}~\bibnamefont {Eriksson}}, \bibinfo {author} {\bibfnamefont
  {L.}~\bibnamefont {Fast}}, \ and\ \bibinfo {author} {\bibfnamefont {J.~M.}\
  \bibnamefont {Wills}},\ }\href {\doibase 10.1103/PhysRevLett.83.3900}
  {\bibfield  {journal} {\bibinfo  {journal} {Phys. Rev. Lett.}\ }\textbf
  {\bibinfo {volume} {83}},\ \bibinfo {pages} {3900} (\bibinfo {year}
  {1999})}\BibitemShut {NoStop}%
\bibitem [{\citenamefont {Jarrige}\ \emph {et~al.}(2013)\citenamefont
  {Jarrige}, \citenamefont {Yamaoka}, \citenamefont {Rueff}, \citenamefont
  {Lin}, \citenamefont {Taguchi}, \citenamefont {Hiraoka}, \citenamefont
  {Ishii}, \citenamefont {Tsuei}, \citenamefont {Imura}, \citenamefont
  {Matsumura}, \citenamefont {Ochiai}, \citenamefont {Suzuki},\ and\
  \citenamefont {Kotani}}]{PhysRevB.87.115107}%
  \BibitemOpen
  \bibfield  {author} {\bibinfo {author} {\bibfnamefont {I.}~\bibnamefont
  {Jarrige}}, \bibinfo {author} {\bibfnamefont {H.}~\bibnamefont {Yamaoka}},
  \bibinfo {author} {\bibfnamefont {J.-P.}\ \bibnamefont {Rueff}}, \bibinfo
  {author} {\bibfnamefont {J.-F.}\ \bibnamefont {Lin}}, \bibinfo {author}
  {\bibfnamefont {M.}~\bibnamefont {Taguchi}}, \bibinfo {author} {\bibfnamefont
  {N.}~\bibnamefont {Hiraoka}}, \bibinfo {author} {\bibfnamefont
  {H.}~\bibnamefont {Ishii}}, \bibinfo {author} {\bibfnamefont {K.~D.}\
  \bibnamefont {Tsuei}}, \bibinfo {author} {\bibfnamefont {K.}~\bibnamefont
  {Imura}}, \bibinfo {author} {\bibfnamefont {T.}~\bibnamefont {Matsumura}},
  \bibinfo {author} {\bibfnamefont {A.}~\bibnamefont {Ochiai}}, \bibinfo
  {author} {\bibfnamefont {H.~S.}\ \bibnamefont {Suzuki}}, \ and\ \bibinfo
  {author} {\bibfnamefont {A.}~\bibnamefont {Kotani}},\ }\href {\doibase
  10.1103/PhysRevB.87.115107} {\bibfield  {journal} {\bibinfo  {journal} {Phys.
  Rev. B}\ }\textbf {\bibinfo {volume} {87}},\ \bibinfo {pages} {115107}
  (\bibinfo {year} {2013})}\BibitemShut {NoStop}%
\bibitem [{\citenamefont {Haule}\ \emph {et~al.}(2005)\citenamefont {Haule},
  \citenamefont {Oudovenko}, \citenamefont {Savrasov},\ and\ \citenamefont
  {Kotliar}}]{PhysRevLett.94.036401}%
  \BibitemOpen
  \bibfield  {author} {\bibinfo {author} {\bibfnamefont {K.}~\bibnamefont
  {Haule}}, \bibinfo {author} {\bibfnamefont {V.}~\bibnamefont {Oudovenko}},
  \bibinfo {author} {\bibfnamefont {S.~Y.}\ \bibnamefont {Savrasov}}, \ and\
  \bibinfo {author} {\bibfnamefont {G.}~\bibnamefont {Kotliar}},\ }\href
  {\doibase 10.1103/PhysRevLett.94.036401} {\bibfield  {journal} {\bibinfo
  {journal} {Phys. Rev. Lett.}\ }\textbf {\bibinfo {volume} {94}},\ \bibinfo
  {pages} {036401} (\bibinfo {year} {2005})}\BibitemShut {NoStop}%
\bibitem [{\citenamefont {Tran}\ \emph {et~al.}(2015)\citenamefont {Tran},
  \citenamefont {Blaha}, \citenamefont {Betzinger},\ and\ \citenamefont
  {Bl\"ugel}}]{PhysRevB.91.165121}%
  \BibitemOpen
  \bibfield  {author} {\bibinfo {author} {\bibfnamefont {F.}~\bibnamefont
  {Tran}}, \bibinfo {author} {\bibfnamefont {P.}~\bibnamefont {Blaha}},
  \bibinfo {author} {\bibfnamefont {M.}~\bibnamefont {Betzinger}}, \ and\
  \bibinfo {author} {\bibfnamefont {S.}~\bibnamefont {Bl\"ugel}},\ }\href
  {\doibase 10.1103/PhysRevB.91.165121} {\bibfield  {journal} {\bibinfo
  {journal} {Phys. Rev. B}\ }\textbf {\bibinfo {volume} {91}},\ \bibinfo
  {pages} {165121} (\bibinfo {year} {2015})}\BibitemShut {NoStop}%
\bibitem [{\citenamefont {Armiento}\ and\ \citenamefont
  {K\"ummel}(2013)}]{PhysRevLett.111.036402}%
  \BibitemOpen
  \bibfield  {author} {\bibinfo {author} {\bibfnamefont {R.}~\bibnamefont
  {Armiento}}\ and\ \bibinfo {author} {\bibfnamefont {S.}~\bibnamefont
  {K\"ummel}},\ }\href {\doibase 10.1103/PhysRevLett.111.036402} {\bibfield
  {journal} {\bibinfo  {journal} {Phys. Rev. Lett.}\ }\textbf {\bibinfo
  {volume} {111}},\ \bibinfo {pages} {036402} (\bibinfo {year}
  {2013})}\BibitemShut {NoStop}%
\bibitem [{\citenamefont {Engel}\ and\ \citenamefont
  {Vosko}(1993)}]{PhysRevB.47.13164}%
  \BibitemOpen
  \bibfield  {author} {\bibinfo {author} {\bibfnamefont {E.}~\bibnamefont
  {Engel}}\ and\ \bibinfo {author} {\bibfnamefont {S.~H.}\ \bibnamefont
  {Vosko}},\ }\href {\doibase 10.1103/PhysRevB.47.13164} {\bibfield  {journal}
  {\bibinfo  {journal} {Phys. Rev. B}\ }\textbf {\bibinfo {volume} {47}},\
  \bibinfo {pages} {13164} (\bibinfo {year} {1993})}\BibitemShut {NoStop}%
\bibitem [{\citenamefont {van Leeuwen}\ and\ \citenamefont
  {Baerends}(1994)}]{PhysRevA.49.2421}%
  \BibitemOpen
  \bibfield  {author} {\bibinfo {author} {\bibfnamefont {R.}~\bibnamefont {van
  Leeuwen}}\ and\ \bibinfo {author} {\bibfnamefont {E.~J.}\ \bibnamefont
  {Baerends}},\ }\href {\doibase 10.1103/PhysRevA.49.2421} {\bibfield
  {journal} {\bibinfo  {journal} {Phys. Rev. A}\ }\textbf {\bibinfo {volume}
  {49}},\ \bibinfo {pages} {2421} (\bibinfo {year} {1994})}\BibitemShut
  {NoStop}%
\end{thebibliography}%

\end{document}